\documentclass[prapplied,amsmath,twocolumn,superscriptaddress]{revtex4}

\usepackage{amsmath,amssymb,graphicx}
\usepackage{ mathrsfs }
\usepackage{xcolor}
\usepackage[makeroom]{cancel}
\usepackage[noend]{algpseudocode}
\usepackage{algorithm}
\usepackage{graphicx}
\usepackage{subfigure}
\usepackage{multirow}
\usepackage{xurl}

\newcommand{\ket}[1]{\left|#1\right\rangle}
\newcommand{\bra}[1]{\left\langle#1\right|}

\newcommand{\avg}[1]{\left\langle#1\right\rangle}

\newcommand{\remove}[1]{}

\begin{document}
\title{
Partitioned Iterative Quantum Scheduling of Satellites for Urgent Disaster Response: Case study of Wildfire}
\author{Lucas~T.~Braydwood}
\email{Lucas.T.Brady@nasa.gov}
\affiliation{Quantum Artificial Intelligence Laboratory, NASA Ames Research Center, Moffett Field, CA 94035, USA}
\author{Taejin Park}
\affiliation{Earth Science Division, NASA Ames Research Center, Moffett Field, CA 94035, USA}
\affiliation{Bay Area Environmental Research Institute, Moffett Field, CA 94035, USA}
\author{Hirofumi Hashimoto}
\affiliation{Earth Science Division, NASA Ames Research Center, Moffett Field, CA 94035, USA}
\affiliation{California State University Monterey Bay, Seaside, CA 93955, USA}
\author{Zoe Gonzalez Izquierdo}
\thanks{Currently at KBR Inc.}
\affiliation{Quantum Artificial Intelligence Laboratory, NASA Ames Research Center, Moffett Field, CA 94035, USA}
\affiliation{USRA Research Institute for Advanced Computer Science (RIACS), Mountain View, CA 94043, USA}
\author{Andrew Michaelis}
\affiliation{NASA Ames Research Center, Moffett Field, CA 94035, USA}
\author{Eleanor Rieffel}
\thanks{Currently at the University of Queensland, Australia.}
\affiliation{Quantum Artificial Intelligence Laboratory, NASA Ames Research Center, Moffett Field, CA 94035, USA}
\author{Shon Grabbe}
\affiliation{Quantum Artificial Intelligence Laboratory, NASA Ames Research Center, Moffett Field, CA 94035, USA}

\date{\today}
\begin{abstract}
The standard in Earth-observation tasks today is having near real-time access to surface images in response to changing conditions. For instance, as urban environments interface more with wildlands and wildfires become less predictable, their tracking with satellite resources becomes essential. This requires the coordination of increasingly large constellations of satellites, giving rise to challenging computational problems. 
With wildfire detection and tracking as a backdrop, we investigate the power of special purpose and novel computing paradigms to tackle the ensuing satellite scheduling problems, making a compelling case for quantum algorithms. We bring quantum scheduling algorithms closer to implementation by examining both the emerging iterative quantum algorithm framework, which comes with analytic guarantees compared to some classical algorithms, and distributed quantum computing methods whose relevance is on the rise as utility-scale problems begin to get solved with quantum computers. Drawing strength from several computing fronts, we develop a distributed/parallelization scheme in conjunction with the quantum algorithm design and apply these techniques to real-world datasets for wildfire detection. While our quantum subprocesses are currently too small to see significant quantum advantage, our results validate the utility of these techniques, and continue forging the path toward distributed quantum computing.
\end{abstract}

\maketitle

\section{Introduction}

Earth science research faces numerous computational challenges such as acquiring, compressing, analyzing, and interpreting massive amounts of data coming from a variety of sources on the ground, in the air, and in the oceans. We expect to see an increase in this variety, and in the flexibility and controllability of its sources in the near future, bringing to the forefront the problem of efficient allocation of sensing and observational resources. Satellites supporting Earth science observations are one such asset, with their number projected to increase from the tens we have today, to hundreds in the not-too-distant future \cite{WILKINSON2024168584}.

As the number of satellites increases, so too will demand for their use.  One notable use case is imaging satellites tasked to collect near-real time information on fires. 
Wildfires can pose devastating risk to communities as well as natural ecosystems, so identifying actionable wildfire event triggers and optimizing data acquisition tasking has great beneficial potential \cite{UNEP}. Globally, wildfires burn 3–5 million km² of land and emit roughly 8 billion tonnes of CO2 annually \cite{derWarf}. These fires have been responsible for at least 2,500 deaths, 10,500 injuries, and the displacement of 175,000 people worldwide since 1990 \cite{Doerr}. The economic costs of wildfires can also be significant. For example, the total economic costs of the California wildfires from 2017 to 2021 were estimated at US\$117.4 billion per year \cite{Paci}. Thus, identifying key wildfire triggers and optimizing data acquisition strategies could offer substantial benefits. However, scheduling these imaging requests from a constellation of satellites is a hard computational task in itself, with even simple cases being equivalent to solving the NP-hard Maximum Independent Set problem.
And as problem complexity increases, it becomes remarkably challenging to arrive at a solution, even with approximate and heuristic methods.

Such difficulty encourages the use of specialized and novel computing schemes, such as distributed quantum computing which we study here.  Quantum optimization \cite{Farhi2000,Farhi2014,Hadfield2019} has a rich history of application to combinatorial optimization tasks both in theory and practice, especially on the Maximum Independent Set problem that underlies scheduling tasks \cite{djidjev2018efficient,yu2021quantum,Ebadi2022,Finzgar2023,Brady2023}.

One of the key problems with current generation quantum optimization problems is that quantum hardware is too small to accommodate utility scale problems.  Even with larger computers on the horizon, quantum computers will need to exist in a networked state if they are to reach these sizes, relying on communication channels between quantum processor units (QPUs).  In this work, we take a promising set of quantum optimization algorithms using iterative problem reduction \cite{Bravyi2020,Bravyi2022,Finzgar2023,Brady2023} and modify it to work in a distributed quantum network environment using solely classical communication channels between QPUs, or a single QPU working in serial.  

While this approach shares similarities with other distributed and divide -and-conquer schemes using quantum computers \cite{Bian2014,Bian2016,Lackey2018}, one major advantage is that it provides the ability to use the novel structure of Iterative Quantum Algorithms \cite{Brady2023} to incorporate the communication channels directly into the algorithm.  Furthermore, this distributed structure can be integrated into classical greedy algorithms for the satellite scheduling problem, opening another path for novel algorithmic study.


We investigate the application of these quantum and classical algorithmic techniques to the problem of allocating observational assets for the acquisition and processing of Earth Science data. In particular, our data represents locations of interest for wildfire detection, and we model a scheduling problem in which one or several satellites are tasked with imaging these locations. 

We model the general scheduling optimization problem, and create a benchmark set of small problem instances based on actionable wildfire triggers filtered from Earth science observations that are suitable for testing quantum or hybrid quantum-classical approaches, as well as application-scale problems whose more efficient solution would have a significant practical impact. Our work both advances classical approaches and explores the potential of quantum computing for the computational challenges in Earth Science.

In the next Section, \ref{sec:fire_monitoring}, we focus on the wildfire monitoring task, describing the general needs in this problem before discussing the datasets used in this study in Section \ref{sec:datasets}.  Section \ref{sec:satellite_sched_pb} covers the formulation of this problem into a Maximum Independent Set framework and then into a form amenable for current quantum computers.  The general framework for the quantum algorithms is developed in Section \ref{sec:IQAs}, and the distributed partitioning scheme is described in Section \ref{sec:distributed}.  Finally, numeric results are discussed in Section \ref{sec:numeric_results} before conclusions and discussions in Section \ref{sec:conclusion}.

\section{Urgent Demand for Near-Real Time Active Fire Monitoring}
\label{sec:fire_monitoring}

There is an urgent demand for near-real-time active fire monitoring due to the increasing frequency and severity of wildfires worldwide. Timely detection and response are crucial for mitigating the devastating impacts of these fires on ecosystems, communities, and infrastructure. 

Advanced near-real-time fire monitoring systems can provide invaluable data and insights to aid in early detection, rapid response, and proactive decision-making, facilitating the efficient allocation of firefighting resources and the implementation of effective evacuation strategies. However, current systematic civilian Earth observing missions using polar-orbiting satellites (e.g., MODIS, VIIRS, Landsat) often lack the capacity for dynamic and high temporal-resolution sampling of unpredictable and short-lived events including wildfires and other natural hazards such as volcanoes and landslides \cite{LI2020111600}. 

Geostationary satellites have the capability to monitor active fire progression in near-real-time \cite{kang2022}, but their coarser spatial resolution often hampers the creation of actionable information for fire fighting, evacuation, and resource allocation \cite{WOOSTER2021112694,LI2020111600,KATO2021102491}. Recent commercial satellite constellations and some civilian missions have provided dynamic tasking capability for capturing urgent and short-lived events. For instance, Planet Lab operates ~20 sub-meter multi-spectral optical sensors onboard SkySats in orbit and offers dynamic tasking service from 5-7 times revisit per day \cite{10281772}. ICEYE has launched 35 X-band SAR satellites so far and provided under 6 hours revisit time for any given location on Earth, ushering in an unprecedented era of Earth observations \cite{9324531}. 

Given the increasing demands for timely and urgent tasking and the proliferation of satellite constellations across private and public sectors, optimizing scheduling and allocating observing capability for dynamic taskings poses a significant challenge. This challenge is particularly pronounced in multi-fire incidents occurring over wildland-urban interface regions, where infrastructure and residential areas are at imminent risk of being impacted.

\section{Wildfire Data Set}
\label{sec:datasets}

This section describes how the benchmark problem set for testing the quantum algorithm was developed (see Sec.~\ref{sec:problem_set}) using near-real time active fire records that are discussed in Sec.~\ref{sec:fire_records}, the Wildland Urban Interface (WUI) \cite{WUI_1990_2020} as discussed in Sec.~\ref{sec:wui}, and satellite orbital tracks as discussed in Sec.~\ref{sec:satellites}.

\subsection{Near-Real Time Active Fire Record}
\label{sec:fire_records}
To provide near real-time active fire information to the public, NOAA has developed the Wild Fire Automated Biomass Burning Algorithm (WFABBA) from Geostationary Operational Environmental Satellites (GOES)-8 \cite{Koltunov2012}.
The WFABBA identifies fire pixels in satellite images and estimates sub-pixel fire characterizations, such as fire size, fire radiative power (FRP), and fire temperature for each fire-detected pixel. The GOES-16 improved the spatiotemporal resolution of the previous GOES series and can detect fire spots in 2 km resolution every 5 minutes for the conterminous US and 10 minutes for GOES full disk domain \cite{Schmit2017}.
Those improvements enabled us to detect active fire locations over the conterminous US by applying WFABBA to GOES-16 in near real-time. We used GOES-16 data for skimming the large area to spot the active fire region, where we will request high-resolution commercial satellite images to capture more details of active fire status.

NOAA GOES-16 Level 2+ Fire/Hot Spot Characterization (FDC) products \cite{Schmidt2020} 
were used to detect and characterize active fire events over the western US.  The FDC products deploy WFABBA which ingest GOES-16 Level 1 data and detect active fire pixels. In general, fires produce a stronger signal in the midwave infrared bands (around 4 microns) than they do in the longwave infrared bands (such as 11 microns). This differential response is a basis for most fire detection and characterization algorithms.

The WFABBA is a multi-spectral thresholding algorithm that uses the shortwave 0.64 µm (ABI Channel 2, available during the daytime) and the 3.9 µm and 11.2 µm bands (Channel 7 and 14) to locate fires and retrieve sub-pixel fire characteristics. The algorithm initially identifies potential active fire candidate pixels from the 3.9 µm and 11.2 µm bands. Then the algorithm incorporates ancillary data to screen for false alarms, correct for water vapor attenuation, surface emissivity, solar reflectivity, and semi-transparent clouds \cite{Schmidt2020}.
The algorithm will characterize fire properties in three ways every 5- or 10-minute: instantaneous fire size, instantaneous fire temperature, and fire radiative power.

\begin{figure}%
    \centering
    \subfigure[]{%
        \label{fig:goes_sample}%
        \includegraphics[height=1.6in]{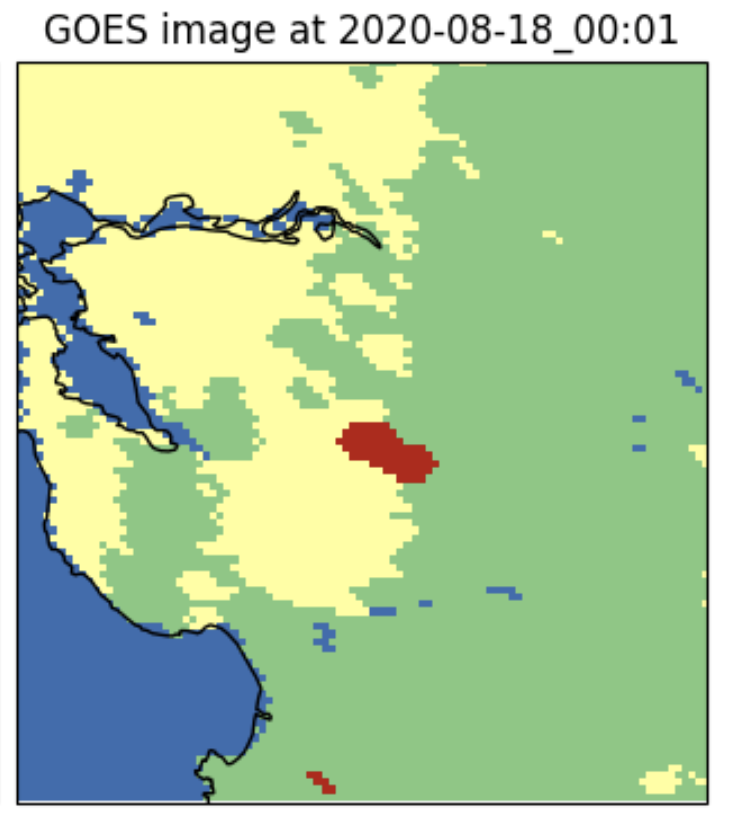}}%
        \qquad
   \subfigure[]{%
        \label{fig:wui_sample}%
        \includegraphics[height=1.5in]{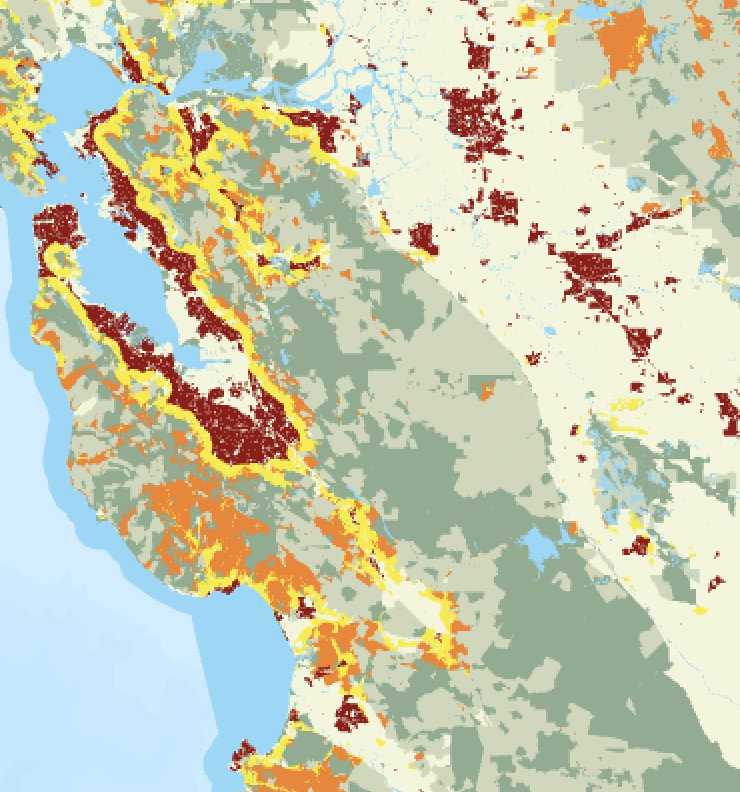}}%
    \caption{(a) Sample GOES-16 FDC imagery from Aug. 18, 2020 at 00:01 UTC. Red denotes an active fire area, green regions with no fire, yellow clouds, and blue water. (b) Corresponding WUI image from the U.S. Forest Service \cite{usfc2023}. Yellow and orange indicate wildland urban interface and intermix, respectively. No housing indicated in dark green and very low housing density shown in medium green. Low and very low housing density in light green, medium and high housing density in maroon, and water indicated in blue.  }
\end{figure}

In this study, we used a 5-minute time step for creating an active fire trigger library. We then utilized the `FireTracks' algorithm (available at https://github.com/dominiktraxl/firetracks) to generate and summarize the characteristics of spatiotemporally tracked fire components. Since each active fire detection record consists of pixel-wise active fire masks, it is necessary to group them and track the progression of individual fires for effective fire monitoring and tasking. The algorithm employs a 3D-Moore neighborhood approach to identify a spatiotemporal cluster as the union of nearest neighbors of fire events in the discrete space-time grid defined by the resolution of the GOES FDC fire masks. We then use the clustered individual fire extent and active fire line information (i.e., perimeter in vector format) to inform the dynamic scheduling systems about event’s location and time to prioritize ongoing disaster events. Data sets were available in five minute increments at a 2 km resolution from July 19, 2020 through Sept. 28, 2020 for this study. An example of the GOES data used is shown in Fig.~\ref{fig:goes_sample}.

\subsection{Wildland Urban Interface (WUI)}
\label{sec:wui}
The WUI refers to a line, area, or zone where human development and infrastructure intersect with or neighbor natural areas prone to wildfires, such as forests, grasslands, or shrublands. This interface is characterized by the close proximity of homes, businesses, or other structures to vegetation \cite{WUI_1990_2020, WUI_2014}. Here, we use WUI to identify the active fire records where additional satellite imagery would be beneficial by prioritizing the fire records in which housing was found to meet or intermix with the undeveloped wildland, see for example Fig.~\ref{fig:wui_sample}. The specific WUI values used are shown in Table~\ref{tab:wui_value}.

\begin{figure}
    \centering
    \includegraphics[width=3.2in]{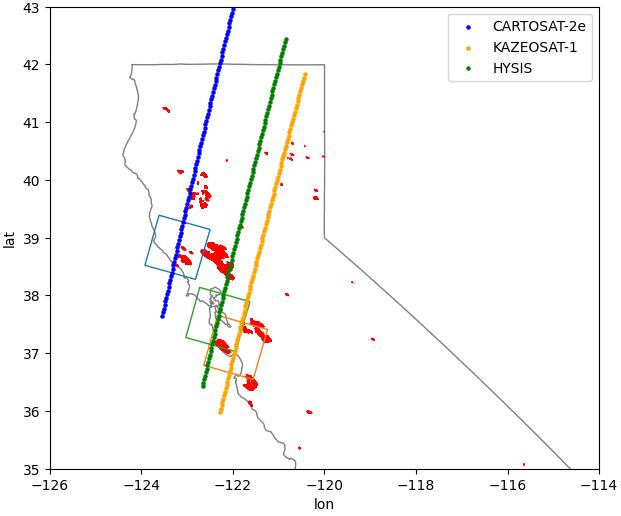}
    \caption{Lines are orbital tracks in one-second increments for CARTOSAT-2e (blue line), KAZEOSAT-1 (yellow line), and HYSIS (green line) on July 25, 2023. Orbital tracks have been synchronized to pass over the CA coast in unison. Red filled polygons are the intersection of GOES FireTracks data from July 19 – Sept 28, 2020 with the WUI low/med/high data. Rectangles are sample, notional field of views (FOVs).}
    \label{fig:firetargets}
\end{figure}

To generate the satellite scheduling problems used in this study, first we determine the set of potential wildfire events to image. This requires calculating the intersection between WUI and the clustered individual fire extent and active fire line information in the form of a perimeter in vector format, as described in Sec.~\ref{sec:fire_records} (see red polygons in Fig.~\ref{fig:firetargets} for an example). How best to use WUI in this context is an active area of research that warrants further study.   


\begin{table*}[]
    \centering
    \caption{WUI values with associated housing density and wildland vegetation considered (see \cite{WUI_1990_2020} for additional details).}
    \label{tab:wui_value}
    \begin{ruledtabular}
        \begin{tabular}{ccc}
            WUI Identifier & Housing Density (units/$km^2$) & Wildland Vegetation\\
            \hline
            Med\_Dens\_Interface & between 49.4 and 741.2  & $\leq~50\%$ and within 2.4 km of area with $\geq 75\%$ \\
            Med\_Dens\_Intermix & between 49.4 and 741.2 & $>50\%$  \\
            High\_Dens\_Intermix & $\geq 741.3$ & $>50\%$ \\
            High\_Dens\_Interface & $\geq 741.3$ & $\leq 50$ and within 2.4 km of area with $\geq 75\%$ \\
            Low\_Dens\_Interface & between 6.2 and 49.2 & $\leq 50\%$ and within 2.4 km of area with $\geq 75\%$ \\
            Low\_Dens\_Intermix & between 6.2 and 49.4 & $>50\%$ \\
            Very\_Low\_Dens\_Veg & $< 6.2$ & $> 50\%$ \\
        \end{tabular}
    \end{ruledtabular}
\end{table*}

\subsection{Satellite Orbital Tracks}
\label{sec:satellites}

The orbital tracks for the set of satellites considered in the satellite scheduling problem (see Sec.~\ref{sec:satellite_sched_pb}) were obtained by running NASA's Navigation and Ancillary Information facility (NAIF) SPICE (Spacecraft Planet Instrument C-matrix Events) framework \cite{ACTON199665, ACTON20189} using two-line elements (TLE) datasets for the CARTOSAT-2e, KAZEOSAT-1, and HYSIS satellites that traversed over Northern California on July 25, 2023. The orbital tracks for these satellites were selected due to their proximity to numerous wildfire sites over Northern California. Note that for this study, the actual sensors onboard these satellites are not considered. Instead, we aim to provide flexibility by considering imaging sensors with variable slew maneuver rates as discussed in Sec.~\ref{sec:satellite_sched_pb}. Additionally, we synchronize the orbital tracks for these three satellites so they cross the California coast at approximately the same time. This way, problem instances with constellations of up to three satellites could be considered.

\subsection{Problem Set Generation}
\label{sec:problem_set}

Given the vectorized perimeters formed by taking the intersection of the WUI and the clustered individual fire extent and active fire line information (see Sec.~\ref{sec:wui})---also referred to as `wildfire targets' below---along with the orbital tracks (see Sec.~\ref{sec:satellites}), problem instances are generated by determining when the `wildfire targets' are within the field of view of the satellites as a function of time. 

For simplicity, the satellite's field of view (FOV), which captures the roll and pitch capabilities of the satellite, was represented as a simple square where larger values were used for more agile satellites and smaller values were used for less agile satellites. In future work, this should be refined to better capture the dynamics of the satellite. Nevertheless, our current approach allowed us to generate problem instance of various sizes for testing the algorithms and for generating the results appearing in Sec.~\ref{sec:numeric_results}, and did not restrict us to using the actual FOVs of the satellites being considered. An example of the FOVs at a specific instance of time for each of our three satellites is represented by the squares in Fig.~\ref{fig:firetargets}. Also, as explained below, the imaging requests for our problem instances are identified by taking the intersection of the FOVs and the red polygons in this figure.

Starting with the orbital tracks from Sec.~\ref{sec:satellites} and the vectorized perimeters formed by taking the intersection of the WUI and the fire data (see above) the steps for generating our problem instances follow:

\begin{enumerate}
    \item Define our FOV set. Here, we have used FOV values of 50km x 50km, 100km x 100km, and 150km x 150km. 
    \item Select the satellite(s) that will be used. Either a single satellite or constellations of up to three satellites can be specified using the satellites listed in Sec.~\ref{sec:satellites}.
    \item Select an unused FOV entry from our set of FOVs.
    \item Set $t$ to be the first timestamp in our satellite orbital tracks.
    \item Select all satellite positions at $t$ and generate the latitude and longitude coordinates of the current FOV centered on the current satellite position.
    \item Determine if any of the vectorized perimeters formed from taking the intersection of our WUI and wildfire data intersect with the current FOV. If yes, store $t$, the satellite name, and vectorized perimeter in our database of imaging requests that will be used for testing the algorithm. See the red polygons residing within the three rectangles in Fig.~\ref{fig:firetargets}, for example.
    \item Repeat Steps (3)-(6) until all FOVs have been processed.
    
\end{enumerate}

\section{Formulation of the Satellite Scheduling Problem}
\label{sec:satellite_sched_pb}

Our formulation of the satellite scheduling problem into a quadratic unconstrained binary optimization problem (QUBO) will closely follow the work of Nag \emph{et al.} \cite{Nag2018} and later follow-ups \cite{Stollenwerk2020,Rainjonneau2023,Makarov2023,Quetschlich2023}.

QUBO problems are equivalent to classical Ising models which are commonly encoded into quantum Hamiltonians and used in analog Hamiltonian quantum algorithms.  For instance, many implementations of algorithms such as Quantum Annealing \cite{Kadowaki1998,Farhi2000} and QAOA \cite{Farhi2014,Hadfield2019} use transverse field Ising models as the default formulation, despite being more generalizable in practice.

The general problem we will examine involves a constellation of $s$ satellites orbiting the Earth that need to image $r$ requested sites on the Earth's surface.  We consider a short time window, both because satellites are near the area to be imaged only during that time period and because it would be inefficient to generate binary variables associated with a too large time range, and discretize time within it.  For each request, $j$, that a satellite, $k$, can fulfill at a specific time, $i$, we create an imaging attempt represented by the binary variable $x_{kji}$, where the indices indicate satellite, request, and time respectively.  These variables will take on the value $1$ if that attempt is in the solution set (i.e. the algorithm proposes to make this attempt) and a value $0$ if it is not in the solution set.  Note that for each $(k,j)$ combination of satellite and request, there will only be a set of times $T_{kj}$ that have attempt variables associated with them.

We also associate a weight with each variable, $w_{kji}$, that equals one if the image is taken under ideal circumstances and decreases as the circumstances become less ideal.  For instance, we take $w_{kji} = 1-\frac{\theta_{kji}}{\pi/2}$ where $\theta_{kji}$ is the angle at which the satellite is taking the image.  This weight is easy to change, and none of our results outside the numerics rely on this value.

The core goal of the problem is to maximize the number of attempts made multiplied by their weights. This will be subject to a few hard constraints.  The first constraint is that we only need one image taken for each request.  We shall enforce this strictly to mean that we are only allowed to take one image of each request.  The second hard constraint is that the satellites need some time to maneuver between requests. This involves the satellite rotating to orient itself towards the request site, and is separate from its orbiting movement. An optimal solution would be one where every request is fulfilled by the satellite that is closest to its location (so that the image is taken straight down, rather than at an angle). All told, we can write this down as a constrained optimization problem described by:

\begin{subequations}
\begin{align}
    \max_{x}& \sum_{k=1}^s\sum_{j=1}^r\sum_{i\in T_{kj}} w_{kji} x_{kji}\\\label{eq:req_con}
    \text{s.t.~~}& \sum_{k=1}^s\sum_{i\in T_{kj}} x_{kji} \leq1,~~~~~\forall~j\in[1,r]\\\label{eq:time_con}
    &x_{kj_1i_1}+x_{kj_2i_2}\leq1,~\text{if}~|i_1-i_2|\leq \iota_{kj_1j_2i_1i_2}
\end{align}
\end{subequations}
where the last condition holds for all satellites and all valid combinations of requests.  The quantity $\iota_{kj_1j_2i_1i_2}$ is determined by the maneuvering time of the satellite as it rotates from one request to another. We assume a constant rotation speed (given as one of the problem inputs), and that the satellite first rotates in the $XY$ plane (change in $\theta$ angle only) and subsequently along the $Z$ axis (change in $\phi$ angle only). These assumptions can easily be replaced to model a different maneuvering pattern based on the realistic movements of the satellites. 

The classical algorithms we use have no need to reference a QUBO formulation, just relying on the graph structure of the problem, and for the hybrid quantum-classical algorithms the classical components will make full use of the original constrained formulation listed above.

The QUBO formulation of the satellite scheduling problem is relatively natural but does rely on turning these hard constraints, Eqs.~(\ref{eq:req_con})~\&~(\ref{eq:time_con}), into soft constraints on the problem.  We use this QUBO formulation in a QAOA algorithm where it is much harder to encode constraints in a hard manner.  Such hard encoding often requires symmetries \cite{Hadfield2019} that are difficult if not impossible to engineer for realistic problems.

The basic optimization over binary variables and their respective weights can be directly translated into a qubit Hamiltonian by taking the binary variables to Pauli-$z$ matrices: $x_*\to (\sigma^{(z)}_*+1)/2$.  The constraints get reformulated into pairwise terms.  The time constraints in Eq.~(\ref{eq:time_con}) are already formulated in this way.
Because these are binary variables, an equivalent to Eq.~(\ref{eq:time_con}) would be $x_{kj_1i_1}x_{kj_2i_2} = 0$ which means that at least one of the variables has to be zero (or equivalently that we cannot have both selected).  This style of constraint can be ported over to Pauli matrices just like before.

The constraints in Eq.~(\ref{eq:req_con}) can actually be reformulated in this pairwise manner as well where we replace this full sum by all possible pairwise combinations of elements in the sums.  Formulated in this way, we can write down terms like $x_{kji}x_{k'ji'}=0$ just like we did with the time constraints.

Before we fully translate this over to a quantum problem, we should pause to realize that this is just the form of weighted Maximum Independent Set (MIS).  MIS is the problem of finding the largest set of nodes in a graph such that there are no edges connecting those nodes, with the weighted version just adding weights to how we score the nodes.  When formulated as pairwise constraints, our time and request constraints can both be interpreted as edges between attempt nodes in a graph.  Then our problem is just one of finding the set of attempts that maximizes our objective, subject to not containing any of those edges.

With this in mind, we can formulate a satellite scheduling problem in terms of MIS and then treat it as a purely mathematical problem.  For simplicity then, we drop the cumbersome indexing we used above and just index all our bits by $i\in[1,n]$ where $n$ is an integer counting all possible attempts.  Then we denote that nodes $i$ and $j$ have an edge between them (are constrained to not both be $1$) with the notation $\langle i,j \rangle$.  Note that we could get multiple constraints giving an edge between two nodes, and in this case, we just count one edge.

Then our classical energy function for MIS (we take a minimization convention, so the ground state corresponds to the maximum independent set) is
\begin{equation}
    E(\vec{x}) = -\sum_{i=1}^n w_i x_i + \lambda \sum_{\langle i,j\rangle} x_i x_j.
\end{equation}
The Lagrange multiplier, $\lambda$, is a tunable parameter that determines how hard the constraint is.  In practice it is usually sufficient to just choose $\lambda$ larger than the reward weights, $w_i$.  Because our $w_i\leq1$, we choose $\lambda = 2$.

This classical energy function is then converted into a quantum Hamiltonian via the transformation $x_i\to (\sigma^{(z)}_i+1)/2$.  Ignoring the physically irrelevant constants, the corresponding quantum Hamiltonian is
\begin{equation}
    \label{eq:C}
    \hat{C} = -\frac{1}{2}\sum_{i=1}^n w_i \sigma^{(z)}_i + \frac{\lambda}{4} \sum_{\langle i,j\rangle} \left(\sigma^{(z)}_i\sigma^{(z)}_j+\sigma^{(z)}_i+\sigma^{(z)}_j\right).
\end{equation}
In the next section, we cover how to use these various problem formulations in quantum and classical algorithms.

\section{Iterative Quantum and Classical Algorithms}
\label{sec:IQAs}

To solve the satellite scheduling problem, we turn to iterative algorithms.  On the classical side, we have greedy algorithms for MIS which are successful \cite{Halldorsson1997}, although not the state of the art for MIS solvers. Because they closely resemble our quantum algorithms \cite{Brady2023} in terms or resources used, they offer a good point of comparison.

\subsection{Classical Greedy Algorithms}

The main classical greedy algorithm we use for Maximum Independent Set is named MIN \cite{Halldorsson1997} because of its selection and reduction rules.

Start with a graph and follow these steps:
\begin{enumerate}
    \item Select the node with the minimum degree $d_i$ (number of edges connected to that node).
    \item Add this node to the solution set, then remove it and all its neighbors from the graph. 
    \item Repeat steps (1)-(2) until the graph is empty.
\end{enumerate}
This method can be modified to a weighted case by replacing degree $d_i$ with weighted degree, $(d_i+1)/w_i$ in the initial ranking.


\subsection{QAOA}

QAOA which stands for either the Quantum Approximate Optimization Algorithm \cite{Farhi2014} or the Quantum Alternating Operator Ansatz \cite{Hadfield2019} is a quantum optimization algorithm that uses alternating applications of a problem and driver Hamiltonian to evolve a system from an easy to prepare state to the ground state of the problem Hamiltonian.

The problem Hamiltonian should have the target state as its ground state and good approximate states as low-lying energy states.  We take $\hat{C}$ from Eq.~(\ref{eq:C}) as our problem Hamiltonian.  The mixer should be an easy to engineer Hamiltonian, with a typical choice being
\begin{equation}
    \hat{B} = -\sum_{i=1}^n \sigma_i^{(x)}.
\end{equation}
The initial state of the system is then the ground state of $\hat{B}$, which in this case is $\ket{\varphi_0}$, the uniform superposition over all $\sigma^{(z)}$ basis states.

QAOA prepares its state through a variational bang-bang form where the two Hamiltonians are applied one after the other for variational lengths of time, $\vec{\gamma}$ and $\vec{\beta}$.  A $p$-depth QAOA state would look like
\begin{equation}
    \ket{\psi(\vec{\gamma}, \vec{\beta})} = \left[\prod_{j=1}^p e^{-i\beta_j \hat{B}} e^{-i\gamma_j \hat{C}}\right]\ket{\varphi_0}.
\end{equation}
The $\vec{\gamma}$ and $\vec{\beta}$ parameters are variational.  The algorithm entails choosing some initial $\vec{\gamma}$ and $\vec{\beta}$ values and then preparing the state $\ket{\psi(\vec{\gamma}, \vec{\beta})}$.  We use this state to measure $\avg{\hat{C}} = \bra{\psi(\vec{\gamma}, \vec{\beta})}\hat{C}\ket{\psi(\vec{\gamma}, \vec{\beta})}$.  Since we are looking for the ground state of $\hat{C}$, varying $\vec{\gamma}$ and $\vec{\beta}$ to minimize $\avg{\hat{C}}$ should get us closer to the desired target state.

For the actual variation of $\vec{\gamma}$ and $\vec{\beta}$, we could use any classical outer loop minimization algorithm.  We will only work with low depth ($p\leq 2$) QAOA, where the exact form of the minimizer doesn't matter much.  For completeness, we used the standard NumPy implementation of Nelder-Mead for this minimization.

This quantum optimization algorithm has found decent popularity in recent years, but as we mentioned in the previous section, it treats the constraints in a soft manner, and so it is better suited for unconstrained problems.

\subsection{Iterative QAOA}

To combat the issues of soft constraints and low circuit depth, we turn to a relatively new class of quantum optimization algorithms, called Iterative Quantum Algorithms \cite{Bravyi2020,Bravyi2022,Patel2022,Bae2023,Brady2023}.

These algorithms take QAOA, Quantum Annealing, or another quantum optimization algorithm as a subroutine to be applied iteratively, using information at each iteration to reduce the size or complexity of the problem for the next one.  Previous work \cite{Brady2023} applied an Iterative version of QAOA to Maximum Independent Set Problems such as the Satellite Scheduling problem considered here.  Our algorithms are the same as the MINQ and MMQ algorithm discussed below.  


MINQ follows a structure similar to that of the classical greedy algorithm outlined above, with a modified and expanded first step:
\begin{enumerate}
    \item Run QAOA on the graph problem, optimizing angles, and calculate the expectation values $\avg{\sigma_i^{(z)}}$ for all the qubits.
    \item Select the node with the maximum expectation value $\avg{\sigma_i^{(z)}}$.
    \item Add this node to the solution set and then remove it and all its neighbors from the graph. 
    \item Repeat steps (1)-(3) until the graph is disconnected, and add all remaining nodes to the solution set.
\end{enumerate}
In Ref.~\cite{Brady2023}, it was shown that this algorithm for unweighted MIS and a $p=1$ QAOA circuit performs exactly the same as classical MIN.  That work also showed some improvement of this algorithm over classical MIN for $p>1$ and weighted problems.

Here, we employ a $p=2$ weighted version of MINQ.  If we could run an entire satellite scheduling problem on a quantum computer or simulator, we would then expect this algorithm to perform better than classical MIN.  Unfortunately, realistic satellite scheduling problems are much larger than can fit on current quantum hardware and simulators.  To address this problem, in the next section, we introduce a divide-and-conquer algorithm for MIS and the satellite scheduling problem that is explicitly tailored to working within the algorithmic framework that MIN and MINQ both share.

The MMQ algorithm is similar to the MINQ algorithm just described except for an expansion of steps (2) and (3) above.  In MMQ, we select the node with the maximum \emph{absolute value} of $\avg{\sigma_i^{(z)}}$ in step (2).  If the selected $\avg{\sigma_i^{(z)}}$ is positive, then proceed with step (3) above, but if it is negative, delete the node from the graph, not adding any nodes to the solution set.  This combines elements of classical MIN and a similar MAX algorithm, while incorporating a quantum selection rule for the elimination choice.

\section{Partitioning MIS Problems}
\label{sec:distributed}

One of the biggest problems with running quantum or hybrid algorithms on current quantum hardware and simulators is that real world problems and datasets are often much larger than can fit on these machines.  Even for moderately small problems, involving a single satellite and $\sim\!\!10$ requests, $400$+ bits would be required to specify them.  One of the main contributions of this work is therefore a methodology for partitioning these problems and incorporating divide-and-conquer techniques into the framework of Iterative Quantum Algorithms.

Although divide-and-conquer schemes exist for quantum algorithms \cite{Bravyi2016,Bian2014,Bian2016,Lackey2018,Piveteau2023}, it remains an understudied field.  Many of these divide-and-conquer schemes are concerned with how to cut small numbers of multiqubit gates in quantum circuit models and simulating the effects of those gates with single qubit gates, many calls to the quantum computer, and classical post-processing \cite{Bravyi2016,Piveteau2023}.  These schemes often have exponential runtimes in the number of cuts made.

The schemes used in Refs.~\cite{Bian2014,Bian2016,Lackey2018} are more similar to what we propose here, being based on modifications to Hamiltonian computing.  However, they perform modifications to the full Hamiltonian based on information obtained from a full run of the algorithm.  On the other hand, we focus on a divide-and-conquer approach tailored to iterative algorithms where we update only small pieces of the problem at any given time.  The partitioning in those references also occurs at the level of the constraints; whereas, we consider partitions directly of the nodes/attempts themselves.

We describe our partitioning problem in terms of the MIS graph, but one of the benefits of the QUBO formulation is that the biases on our qubits will retain a memory of how many edges they had, even after some of those edges have been cut and no longer appear in the graph.  Below we describe our process in detail, both applied to the classical greedy algorithm and the quantum MINQ algorithm.

\subsection{Graph Partitioning}

Optimal Graph Partitioning takes in a graph with nodes and edges and seeks to find the partitioning of those nodes into $k$ roughly equal sets of nodes such that the number of edges crossing between sets is minimized.  This problem is in general NP-Hard, but many good and efficient approximate algorithms exist \cite{KaHIP}.  For the system sizes we consider, these open source packages are remarkably good, and in fact can be used for graphs on the order of social network graphs.

We use an open source graph partitioning package, KaHIP \cite{KaHIP}, specifically, the KaFFPa algorithm, set to FAST mode and with uniform vertex and edge weightings.  These choices are rather simplistic, but we are cognizant that much of the power here could come from the graph partitioning, so we rely on the exact same partitioning for both the quantum and classical methods and focus on pure speed over quality in the graph partitioning.

The main inputs that go into the graph partitioning algorithm are the number of partitions $k$, and the allowable imbalance, $\epsilon$.  This parameter determines the size differences allowed among partitions; specifically, the sizes of the partitions, $\{n_i\}$, must satisfy $n_i\leq (1+\epsilon)n/k \forall i$, with $n$ being the total number of nodes.  Furthermore, since our limitations are based on number of qubits for our simulators, we choose $k$ based off the approximate maximum number of qubits, $m$, allowed in our system, up to the imbalance factor.  So we choose $k = \lceil\frac{n}{m}\rceil$.  In our simulations we report the $m$ and $\epsilon$ values used.

Once a partitioning of the problem is decided, we split the problem into those pieces, removing the edges that were cut.  For the classical greedy algorithm, the partitioned graphs have no lingering connections to each other, and they are ready to be used at this point.  For the quantum algorithms, the QUBO problem includes evidence about the edges in both the connectivity matrix and the biases (an artifact from switching from 0/1 for MIS to $\pm1$ for the qubits).  Our code necessarily cuts the graph edges, but we then have an option to remove the evidence from the biases which are continuing to penalize each bit based off neighbors in other sub-graphs.  In practice, we leave this bias information in place since it reflects how many edges each bit actually has.

\subsection{Recombination}

On each sub-graph we then perform iterative QAOA or a classical greedy algorithm for one step to identify which node in that sub-graph is going to be fixed.  Before we actually fix that node though, we need to communicate between the nodes to (a) see what other nodes in other graphs can also be eliminated via logical inference based on the independence constraints of the problem and (b) see if any nodes we are about to fix to be in the set share a cut edge to a node in another sub-graph that has also been selected to be in the set.  In this second case, we then need to determine which if either of these nodes we are going to actually fix to be in the set.

Breaking conflicts and logical inference across different sub-graphs therefore make up the majority of the work we do in this recombination step.  There are multiple ways that conflicts could be decided.  The simplest would be to ignore both the conflicting reduction rules and recombine only with reductions that were not in conflict.  Unfortunately, if such conflicts are never actually broken, it is possible for the algorithm to get into an infinite loop.  Instead, what we do is go back to the metric used in the selection criteria, the expectation value of the qubit in QAOA or the degree of the graph in classical greedy, and break the conflict in favor of accepting the reduction step stemming from the better selection criteria.  In the event that there is still a tie (unlikely for the quantum algorithm with continuous selection criteria but likely for greedy with discrete criteria), we break the tie arbitrarily.

Once all conflicts have been resolved and all logical inferences across sub-graphs have been carried out, the previously cut edges that still exist are restored, and the graph is recombined.  This concludes one iteration of our algorithm.  The entire graph partitioning is then repeated with this new graph.

\subsection{Sparsity Concerns}

In principle repartitioning the graph at every iteration is a costly procedure that would ideally be avoided.  Instead it would be more efficient to keep the original graph partitioning for multiple iterations.  Unfortunately, given the nature of our problem, this is unfeasible.

Given how we construct our graphs, they tend to consist of several highly connected sub-components, connected to each other by dense but varying webs of connections.  This means that our graphs tend to be quite dense overall, a feature that would diminish if our simulators could look at large enough system sizes to consider longer time-horizons.  But as it is, when we fix one node to be in the set, logical inference allows us to immediately fix a large number of other nodes to be not in the set.  This reduces down the problem size quickly but also means that often a single reduction step can eliminate a large fraction of the variables in a sub-graph.  Therefore, often what is left in a sub-graph is trivial at that point, necessitating a full recombination of the graph to get access to problems that are still interesting to solve.

In general, the density of our graphs is a big detriment to our numeric results and is the primary reason why we think the quantum algorithm results are not favorable when compared to classical results.  While this could be eased by larger simulators and access to larger actual quantum hardware, we also take this as a challenge for the future development of more advanced and quantum specific partitioning means.  In many ways, the divide and conquer schemes employed in this paper are naive first steps that should be taken as something to refine and hone in future studies.

\section{Numeric Results}
\label{sec:numeric_results}

\begin{figure}
    \centering
    \subfigure[~KAZEOSAT]{\includegraphics[width=0.49\textwidth]{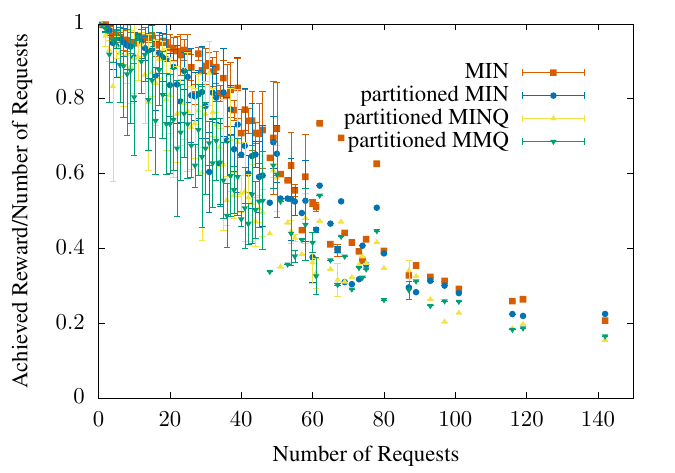}}
    \subfigure[~CARTOSAT]{\includegraphics[width=0.49\textwidth]{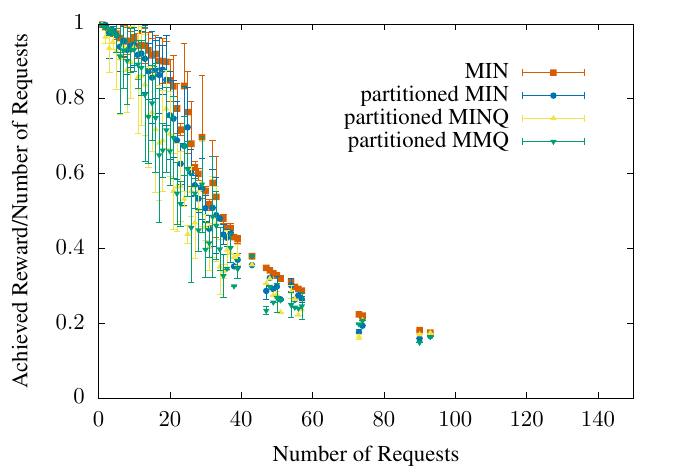}}
    \subfigure[~HYSIS]{\includegraphics[width=0.49\textwidth]{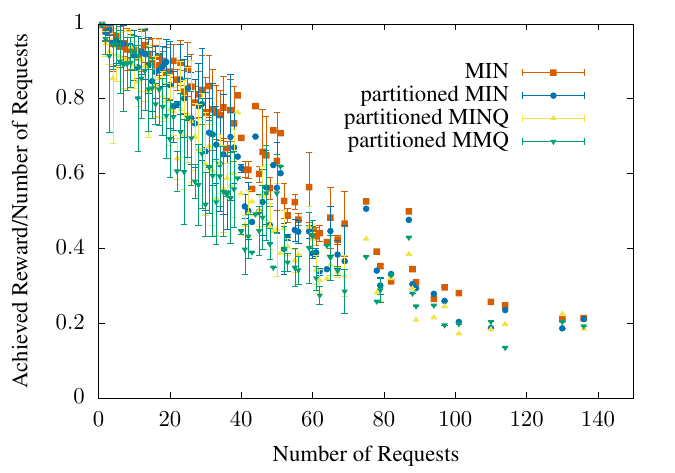}}
    \caption{Simulation results for three different satellites, using the classical algorithms MIN and partitioned MIN as well as two quantum algorithms, both partitioned.  The $y$ axis represents the achieved reward of the algorithm divided by the number of requests, with a value of $1$ corresponding to satisfying every request from an optimal angle.  The data is averaged over all instances for the indicated satellites with higher request simulations coming from longer time bins.}
    \label{fig:KAZEOSAT}
\end{figure}

The three satellites used in this study KAZEOSAT, HYSIS, and CARTOSAT were all used over a set period of time as described above, with imaging requests possible at every discrete time step.  In order to create more simulations, we considered each satellite separately in a given time window or bin and then asked the scheduling algorithms to optimize the satellite's jobs for that time period.

For each satellite, we had finite ticks of time, labeled with numbers ranging $[815,859]$.  To create a time bin, we took a range here of sequential time ticks, with length $1$, $2$, $4$, $6$, and $8$.  The time bins were chosen to overlap with half of the previous time bin.  We did this for each satellite, giving a range of larger and smaller simulations, each with a different number of requests, based on what imaging targets were available to that satellite in the chosen time window.

Furthermore, these simulations were further expanded by considering different levels of WUI data, considering normal interface levels, all the way down to low and very low levels, with simulations covering all interface requests from that level and above.  Thus for each time bin and satellite, three additional scheduling tasks were performed---one for each of these WUI levels.

Results of these simulations are shown in Fig.~\ref{fig:KAZEOSAT}.  The number of requests is shown on the $x$ axis. When there are several different scenarios with the same number of requests (which happens often at lower values), the data point corresponds to the average over all such scenarios, with error bars representing the standard deviation about the mean. As the number of requests increases, this is no longer the case, so no error bars are shown.  
The $y$-axis shows the reward generated divided by the number of requests made.  If the algorithm manages to schedule every request from an optimal angle, then this ratio is $1$.  As the number of requests increases, the likelihood of a perfect schedule being attainable decreases, so the decline seen on the plots is expected and reflects the nature of the scheduling task rather than the quality of the solvers.

As can be seen, the quantum algorithms are not competitive with the classical solvers, even considering the partitioned classical solver.  We attribute this to the size of the partitions being solved, necessitated by the number of qubits allowed by our quantum simulation software (we restricted to an average of 18 qubits per partition).  This partitioning is too small to accurately collect necessary graph features.  We expect the performance of the quantum algorithm to improve as partition sizes become larger.  Analytics for all system sizes and numerics for small system sizes \cite{Brady2023}, indicate that in the limit of no partitions, the quantum algorithms should outperform their classical counterparts.  It is also possible that improvements in the partitioning scheme could lead to better results relative to the partitioned greedy algorithm.
The gap in performance also shrinks as system size and number of requests increases, indicating again that our results might be unduly influenced by too small of systems.

\section{Conclusion}
\label{sec:conclusion}

We provide details of our wildfire dataset and methodology, addressing the ongoing need within the community for solutions to low-latency (geostationary satellite-based fire event detection), high-spatial-resolution (dynamic tasking) mapping of active fire and its spread (e.g., fire frontline). These factors are recognized as crucial for rapidly responding to fire impacts, efficiently allocating resources, and planning for subsequent recovery. The timely data obtained through demonstrated near-real-time event detection, coupled with dynamic tasking capabilities, holds significant promise for enhancing our ability to collect data in response to short-term and unpredictable events that affect the Earth system and global community. The dynamic tasking strategy tested here for wildfires can be applied to any other disaster responses and urgent events requiring rapid tasking, including volcanic eruptions, flooding, and landslides.

On the quantum algorithm front, we took the emerging field of iterative quantum algorithms and combined it with methods from divide and conquer and distributed computing.  As quantum computers seek to tackle utility-scale problems, distributing tasks over multiple QPUs, even in the fault tolerant regime, is going to be necessary.  Understanding how to properly interconnect multiple processes together and recombine results is a key challenge facing quantum computing.  This work and methodology is important for early applications of such distributed quantum computing because it relies solely on classical communication between different QPUs, allowing distribution without needing to wait for the development of effective quantum communication channels.

Our results do not show a major advantage of quantum algorithms in this scheme, but we strongly suspect that this lack of relative performance is due in large part to the size of the quantum subprocesses.  We were restricted to relatively small simulated QPUs, given the limitations of classical simulators, and while modern quantum hardware could support larger system sizes, the noise levels in such devices often defeat the purpose of trying to solve realistic problems.  With larger subprocesses, a single one can capture more relevant problem features, and we expect these divide and conquer methods to be useful for near-term distributed quantum algorithms.

Of course, future quantum distributed algorithms will need to rely on quantum interconnects between processors.  Such interconnects will often not lead to an exact solution, but the mixture of more advanced parallelization methods with heuristic quantum algorithms provides an intriguing path, especially in the early days of quantum computers reaching small utility-scale problems.

\acknowledgements
We are grateful for support from NASA Ames Research Center, NASA Earth Science Technology Office (ESTO), and DARPA under IAA 8839, Annex 114. ZGI was supported by NASA Academic Mission Services (NAMS), contract number NNA16BD14C.

\bibliography{refs}

\end{document}